\documentclass[11pt]{article}
\pdfoutput=1
\usepackage{jheppub}
\usepackage{blindtext}

\usepackage{mathrsfs}
\usepackage{bm}
\usepackage{paralist}
\usepackage{url}
\usepackage[]{microtype}

\usepackage{tikz-cd}
\usepackage{tikz}
\usepackage{pict2e}
\usepackage{float}
\usepackage{stmaryrd}
\usepackage{geometry}
\usepackage{tensor}
\usepackage{mathtools}

\usepackage{amsmath,amssymb,amsthm,amscd,graphicx}
\usepackage{psfrag}
\usepackage[utf8]{inputenc}
\usepackage{graphicx}
\usepackage{xcolor}
\usepackage{hyperref}
\hypersetup{colorlinks,allcolors=blue}
\definecolor{frangreen}{rgb}{0.040, 0.475, 0.435}

\usepackage[T1]{fontenc}

\usepackage[utf8]{inputenc}
\usepackage{graphicx}
\usepackage{amsmath,amsthm,amssymb,physics,mathtools}
\usepackage{tikz,tikz-cd, dynkin-diagrams}
\usepackage{epigraph}
\usepackage{hhline}
\usepackage{ytableau}
\usepackage{natbib}
\usetikzlibrary{positioning,arrows}
\usetikzlibrary{decorations.pathmorphing}
\usetikzlibrary{decorations.markings}
\usetikzlibrary{matrix}
\usepackage{bbold}

\usepackage{environ}         % provides \BODY
\usepackage{etoolbox}        % provides \ifdimcomp
\usepackage{graphicx}        % provides \resizebox
\usepackage{nomencl}
\makenomenclature

 \usepackage{etoolbox}
\renewcommand\nomgroup[1]{%
  \item[\bfseries
  \ifstrequal{#1}{C}{CFT symbols}{%
  \ifstrequal{#1}{H}{Heun symbols}{%
  \ifstrequal{#1}{F}{CFT symbols - semiclassics}{}}}%
]}
	\newlength{\myl}
\let\origequation=\equation
\let\origendequation=\endequation

\RenewEnviron{equation}{
  \settowidth{\myl}{$\BODY$}                       % calculate width and save as \myl
  \origequation
  \ifdimcomp{\the\linewidth}{>}{\the\myl}
  {\ensuremath{\BODY}}                             % True
  {\resizebox{\linewidth}{!}{\ensuremath{\BODY}}}  % False
  \origendequation
}

\theoremstyle{definition}

\newcommand{\eqlb}[2]{\begin{equation} \label{#1} #2 \end{equation}}
\newcommand{\eq}[1]{\begin{equation} #1 \end{equation}}

\newcommand{\brc}[1]{\left(#1\right)}
\newcommand{\bsq}[1]{\left[#1\right]}

\newcommand{\rme}{\textrm{e}}
\newcommand{\be}{\begin{equation}}
\newcommand{\ee}{\end{equation}}
\newcommand{\ba}{\begin{aligned}}
\newcommand{\ea}{\end{aligned}}
\newcommand{\ben}{\begin{eqnarray}\displaystyle}
\newcommand{\een}{\end{eqnarray}}

\makeatletter
\gdef\@fpheader{}
\makeatother

\title{Black hole scattering amplitudes via analytic small-frequency expansion and monodromy}

\author[a,b]{Gleb Aminov}
\author[c,d,e]{Paolo Arnaudo}

\affiliation[a]{ C.~N. Yang Institute for Theoretical Physics, State University of New York, Stony Brook, NY 11794-3840, USA}
\affiliation[b]{ Simons Center for Geometry and Physics, State University of New York, Stony Brook, NY 11794-3636, USA}
\affiliation[c]{International School of Advanced Studies (SISSA), via Bonomea 265, 34136 Trieste, Italy}
\affiliation[d]{INFN Sezione di Trieste, via Valerio 2, 34127 Trieste, Italy}
\affiliation[e]{Institute for Geometry and Physics, IGAP, via Beirut 2, 34151 Trieste, Italy}

\emailAdd{gleb.aminov@stonybrook.edu}
\emailAdd{parnaudo@sissa.it}
		
\date{\today}

\abstract{We utilize three complementary approaches to pinpoint the exact form of scattering amplitudes in Schwarzschild spacetime. First, we solve the Regge-Wheeler equation perturbatively in the small-frequency regime. We use the obtained solutions to determine the monodromy in the near-spatial infinity region, which leads to a specific partial differential equation on the elements of the scattering matrix. As a result, it can be written in terms of the elements of the infinitesimal generator of the monodromy transformation and an integration constant. This constant is further related to the Nekrasov-Shatashvili free energy through the resummation of infinitely many instantons. The quasinormal mode frequencies are also studied in the small-frequency approximation.
}

\begin{document}

\maketitle

\section{Introduction}
We study linear perturbations around ordinary Schwarzschild black hole (BH) in four dimensions. The class of perturbations we consider is described by a Regge-Wheeler differential equation encoding massless scalar, electromagnetic, and odd gravitational perturbation fields \cite{PhysRev.108.1063}. This can be conveniently rewritten as a $SU(2)$ Seiberg-Witten quantum curve with three fundamental hypermultiplets ($N_f= 3$) \cite{Seiberg:1994rs, Seiberg:1994aj} or, equivalently, as a confluent Heun equation \cite{ronveaux1995heun}. The main goal of this work is to find an analytic description of scattering amplitudes for each type of perturbation field.  To do so, we utilize three complementary approaches: analytic small-frequency expansion, the study of the monodromy properties, and the correspondence with the Seiberg-Witten theory \cite{Aminov:2020yma}. While the first approach is novel, the monodromy and the BH-Seiberg-Witten correspondence has been analyzed from several perspectives, such as supersymmetric gauge theory and Liouville conformal field theory \cite{agt,Piatek:2017fyn,Bonelli:2021uvf,Bianchi:2021xpr,Bianchi:2021mft,Bonelli:2022ten,Dodelson:2022yvn,Consoli:2022eey,Giusto:2023awo,Lei:2023mqx}, monodromy and Painlevé equations \cite{Castro:2013kea, Castro:2013lba,novaes, novaes2014,CarneirodaCunha:2015hzd,   CarneirodaCunha:2019tia,Amado:2020zsr,BarraganAmado:2021uyw,daCunha:2022ewy}, and thermodynamic Bethe ansatz \cite{Fioravanti:2021dce, Gregori:2022xks, Fioravanti:2023zgi}.

To solve the Regge-Wheeler equation, we generalize the perturbative approach to differential equations with regular singular points described in \cite{Aminov:2023jve}. Since the Regge-Wheeler equation has one irregular singular point, we introduce a new set of special functions called \emph{multiple polyexponential integrals} \cite{Aminov:2024aan}. The small parameter, which we denote with $t$, is proportional to the frequency $\omega$ of the perturbation field, leading to a small-frequency expansion of the spectral problem. 
Around the black hole horizon, we select the local solution representing the purely incoming wave. Near spatial infinity, the local solution is represented as a linear combination of the purely outgoing and purely incoming waves:
\eq{\psi_{(\infty)} = \mathcal{A}\,\psi^{\mathrm{out}}_{(\infty)}+\mathcal{B}\,\psi^{\mathrm{in}}_{(\infty)},}
where $\mathcal{A}$ and $\mathcal{B}$ are the reflection and incidence coefficients, respectively.
The elements of the scattering matrix $S\equiv S_{\ell,s}$ are then defined as
\eq{S=\frac{\mathcal{A}}{\mathcal{B}},}
which we will call \emph{scattering elements} for brevity.
The corresponding scattering amplitude is
\begin{equation}
f_s(\omega,\theta)=\frac{1}{2\,\imath\,\omega}\sum_{\ell=0}^{\infty}\brc{2\,\ell+1}\brc{S_{\ell,s}(\omega)-1}P_{\ell}\brc{\cos\theta}.
\end{equation}
The scattering elements $S_{\ell,s}$ have been studied semi-analytically \cite{Futterman_Handler_Matzner_1988,Andersson:1994rk,Mano:1996mf,Mano:1996gn,Mano:1996vt,Decanini:2002ha,Folacci:2019cmc,Ivanov:2022qqt,PhysRevD.108.044050,Saketh:2023bul,Ivanov:2024sds} and, most recently, via gauge theory methods \cite{Bautista:2023sdf}. Closely related are the studies of Green functions in Schwarzschild spacetime \cite{PhysRevD.34.384,Casals:2013mpa,Casals:2015nja,Kyutoku:2022gbr}.

The monodromy of the two solutions $\psi^{\mathrm{out}}_{(\infty)}$ and $\psi^{\mathrm{in}}_{(\infty)}$ around spatial infinity provides us with a partial differential equation on the scattering elements. Solving this equation, we find
the exact $\log t$ dependence of $S$. The remaining integration constant $\phi(t)$ is given by its Taylor expansion in $t$ and can be further related to the Seiberg Witten B-period \cite{nekrasov2003,no2,nekrasov2010,mirmor,mirmor2,Zenkevich2011, Bonelli:2011na}. We perform explicit checks of this relation and specify the resummations that occur to express the instanton expansions as small-frequency expansions. As a result, an alternative expression of the scattering amplitudes in terms of the Nekrasov-Shatashvilli free energy \cite{nekrasov2010} is obtained, which aligns with the recent studies of the Kerr-Compton amplitudes carried out in \cite{Bautista:2023sdf}. Yet another description of the scattering elements was obtained earlier in \cite{Ivanov:2022qqt,Saketh:2023bul,Ivanov:2024sds} using the MST method \cite{Mano:1996mf,Mano:1996gn,Mano:1996vt}. All three approaches are related: the MST method might be "in-between" the instanton and the pure small-frequency approaches. The exact relation between the above approaches is beyond the scope of this paper.

The poles of the scattering elements correspond to the quasinormal mode (QNM) frequencies $\omega_{n,\ell,s}$, which have been extensively studied in recent years \cite{PhysRevD.30.295,PhysRevD.35.3632,Cardoso:2003vt,Konoplya:2004ip,Berti:2009kk,PhysRevD.86.024021,Novaes:2018fry,Hatsuda:2020sbn,Hatsuda:2020egs,Amado:2021erf,Hatsuda:2021gtn,Imaizumi:2022qbi,Imaizumi:2022dgj,Bianchi:2022qph,Dodelson:2023vrw,Bianchi:2023rlt,Bianchi:2023sfs,Hatsuda:2023geo,Arnaudo:2024rhv} due to their relevance for gravitational wave astronomy \cite{PhysRevLett.116.061102}.
The advantage of the small-frequency approach to computing QNM frequencies is that it requires a comparatively small input, two Taylor expansions in $t$:
\eq{\beta_{\ell,s}=\sum_{k\geq 1} \beta_{\ell,s,k} \, t^{2k},\quad 
\varphi_{\ell,s} = \sum_{k\geq 0} \varphi_{\ell,s,k} \,t^{2\,k}.}
For given quantum numbers $\ell$ and $s$, the coefficients $\beta_{\ell,s,k}$ and $\varphi_{\ell,s,k}$ are rational numbers, except for $\varphi_{0,0,0} = \log\brc{7/9}$.
By computing the first $7$ coefficients in the expansion of $\beta_{\ell,s}$, we get the numerical values of QNMs up to $4$ or $5$ digits. On the downside, the radius of convergence for each of these Taylor expansions seems only to capture the fundamental frequency $\omega_{0,\ell,s}$.
By using the NS free energy formulation of the scattering elements, one can rewrite $\beta_{\ell,s}$ and $\varphi_{\ell,s}$ as instanton expansions with two artificially separated parameters $\Lambda \equiv t$ and $\omega$. This will increase the radius of convergence and allow one to compute frequencies with a higher overtone number, $n>0$ \cite{Aminov:2020yma}.

This paper is structured as follows. In Sec.~\ref{sec:method}, we describe the perturbative method and the other techniques used. In Sec.~\ref{sec:RW}, we introduce the three local regions in which the Regge-Wheeler equation is solved and the corresponding leading order solutions. In Sec.~\ref{sec:wavesol}, the higher-order corrections to the solutions are described in terms of the special functions, such as multiple polylogarithms and multiple polyexponential integrals. In Sec.~\ref{sec:monodromy}, we study the monodromy of the two local solutions in the near-spatial infinity region and obtain an explicit expression for the infinitesimal generator of the monodromy matrix. In Sec.~\ref{sec:amplitudes}, the exact dependence of the scattering elements on $\log t$ is determined by solving a specific partial differential equation, and the connection with the Seiberg Witten quantum periods is discussed. In Sec.~\ref{sec:QNM}, we solve for the poles of the scattering elements, finding the fundamental QNM frequencies for different values of quantum numbers $\ell$ and $s$. 
%\textcolor{red}{In Sec.~\ref{sec:conclusion}, we draw our conclusion and discuss possible generalizations and open problems related to the present work.}
In Appendix~\ref{appendix:beta}, the results for the Taylor expansions of $\beta_{\ell,s}$ and $\varphi_{\ell,s}$ are presented. Also, the generic formulas for $\beta_{\ell,s,k}$, $\ell\geq k$ and $\varphi_{\ell,s,k}$, $\ell>k$ are derived from the instanton expansion in Appendix~\ref{appendix:beta}.

\vskip 0.5cm
\noindent {\large {\bf Acknowledgments}}
\vskip 0.5cm
We thank G. Bonelli, M. Dodelson, P. Gavrylenko, A. Grassi, A. Grekov, C. Iossa, Z. Komargodski and A. Tanzini for their valuable remarks and discussions.   

The research of P.A.\ is partly supported by the INFN Iniziativa Specifica GAST and
by the MIUR PRIN Grant 2020KR4KN2 
"String Theory as a bridge between Gauge Theories and Quantum Gravity".
P.A. acknowledges funding from the EU project  
Caligola HORIZON-MSCA-2021-SE-01), Project ID: 101086123.

\section{Outline of the method}\label{sec:method}

We study linear perturbations around the Schwarzschild black hole of mass $M$ that are described by the Regge-Wheeler equation
\begin{equation}\label{eq:ReggeWheeler}
\left(\partial^2_r + \frac{f'(r)}{f(r)} \partial_r +\frac{\omega^2-V(r)}{f(r)^2}\right)\Phi(r)=0,
\end{equation}
where $\rme^{- \imath\, \omega\, t}\,\Phi(r)$ is the radial part of the perturbation and the 
potential is
\begin{equation}
V(r)= f(r)\left[\frac{\ell(\ell+1)}{r^2}+(1-s^2)\left(\frac{2M}{r^3}\right)\right],\quad f(r)=1-\frac{2M}{r}.
\end{equation}
Different values of $s$ correspond to massless scalar ($s=0$), electromagnetic ($s=1$), and odd gravitational ($s=2$) perturbations.

%{\color{red} The QNMs are obtained by imposing particular boundary conditions such that there is no incoming wave at spatial infinity and only an incoming wave is present at the horizon:
%\begin{equation}
%\Phi\brc{r} \sim \left\{
%\begin{array}{ll}
%  \rme^{-\imath\, \omega\, r^{*}}, & r\rightarrow 2 M, \\
%  \rme^{\,\imath\, \omega\, r^{*}}, & r\rightarrow \infty,
%\end{array}
%\right.
%\end{equation}
%where the tortoise coordinate is defined as $r^{*}=r+2\,M \log\brc{r-2\,M}$.}

The starting point of our approach is to rewrite the Regge-Wheeler  equation \eqref{eq:ReggeWheeler} in the form of a quantum spectral curve
\eqlb{eq:Nf3_curve}{\widehat{H}\, \Psi\brc{x}= E\, \Psi\brc{x}}
with the following Hamiltonian:
\eqlb{eq:Nf3_Ham}{\widehat{H}= \hat{p}^2+t^{1/2} \rme^{x} \prod_{j=1}^{2}\brc{\imath\, \hat{p} +M_j +\frac{\hbar}{2}}
 +t^{1/2} \rme^{-x}\brc{\imath\, \hat{p} +M_3 -\frac{\hbar}{2}}.}
After fixing $\hbar=1$, the spectral curve parameters such as energy, masses, and the coupling constant can be written in terms of the black hole parameters $\ell$, $s$, and $M\omega$:
 \eqlb{eq:SW_BH_dict}{E=-\brc{\ell+\frac12}^2,\quad M_1=-s,\quad M_2=s,\quad M_3=0, \quad t =- 4\,\imath \,M \omega.}

The next step is to solve the resulting second-order ODE perturbatively in the small parameter $t$. Following the multi-polylog approach described in \cite{Aminov:2023jve}, we introduce three complex local variables:
\eq{z_R= \frac{2\,M}{r},\quad z_M=\frac{t^{1/2}}{z_R},\quad z_L=\frac{t}{z_R},}
where $z_R$ and $z_L$ are local variables near the horizon and spatial infinity. The intermediate local variable $z_M$ is introduced to improve the efficiency of the perturbation theory. In the near-horizon and near-spatial infinity regions, the wave function can be written as the following Taylor expansion:
\begin{equation}\label{eq:psi_exp0}
\psi(z)=\psi_0(z)+\sum_{k\geq 1} \psi_k(z) t^{k}.
\end{equation}
The same Ansatz works in the intermediate region, except the expansion parameter is $t^{1/2}$ instead of $t$. 
In each region, the leading order solution $\psi_0(z)$ is determined by the boundary condition or the continuity with the adjacent region.
Then, the higher-order corrections $\psi_k(z)$ can be found by solving the following non-homogeneous differential equation:
\begin{equation}\label{eqp}
\psi''_k(z)+ \varphi(z) \psi'_k(z)+ \nu(z) \psi_k(z) +\eta_{k}(z)=0,
\end{equation}
where $\varphi(z)$ and $\nu(z)$ are the leading orders of the corresponding coefficients of the differential equation in the considered region, and the non-homogeneous part $\eta_{k}(z)$ is completely determined in terms of the previous order corrections $\psi_0(z),\dots,\psi_{k-1}(z)$. Using the method of variations of parameters, the corrections $\psi_k(z)$, $k\ge 1$, can be written as
\begin{equation}\label{psicorrections}
\psi_k(z)= c_{k} g_{0}(z) -g_0(z) \int^{z} f_{0}(z')\, \frac{\eta_{k}(z')}{W_0(z')}\, \mathrm{d} z' +f_{0}(z) \int^{z} g_0(z')\, \frac{ \eta_{k}(z')}{W_0(z')}\, \mathrm{d} z',
\end{equation}
where $f_0(z)$ and $g_0(z)$ are the two solutions of the homogeneous part of (\ref{eqp}) and $W_0(z)$ is their Wronskian
\begin{equation}
W_0(z)\equiv f_{0}(z)g'_0(z) - f'_{0}(z)g_0(z).
\end{equation}
Moreover, the solution $f_0(z)$ is chosen such that in the near-horizon region $\psi^{\text{hor}}_0(z)=f_0(z)$ and in the near-spatial infinity region $\psi^{(\infty)}_0\brc{z}=f_0\brc{z}-g_0\brc{z}$. Here, the leading order wave function in the near-spatial infinity region is chosen such that $\psi^{(\infty)}_0\brc{0}=0$.
The integration constants $c_k$ are fixed by imposing the boundary condition or gluing the local solutions.

We start with the region near the horizon and fix the integration constants $c_k$ such that only the incoming wave is present in the wave function:
\eq{\psi_{\text{hor}}=\psi_{\text{hor}}^{\mathrm{in}}.}
After gluing the local solutions, we arrive at the following expression for the wave function in the near-spatial infinity region:
\eq{\label{linearcombinationinfty}
\psi_{(\infty)} = \mathcal{A}\,\psi^{\mathrm{out}}_{(\infty)}+\mathcal{B}\,\psi^{\mathrm{in}}_{(\infty)}.}
The scattering elements $S\equiv S_{\ell,s}$ are defined as the ratio $\mathcal{A}/\mathcal{B}$. They are functions of the $M\cdot\omega$ product (or the parameter $t$ in our case). We will also work with the inverse
\eq{S_{\ell,s}^{-1}=\mathcal{B}/\mathcal{A},}
whose zeroes are the QNM frequencies. In agreement with the previous observations \cite{Casals:2013mpa, Casals:2015nja}, $S$ depends on $\log\brc{t}$ in the small-frequency expansion:
\eq{S=S\brc{t,\log t}.}
The presence of the logarithm function provokes a question: what happens to the scattering elements when we change the branches of the logarithm? The shift in the logarithm $\log t \rightarrow \log t + y$ might change the scattering elements:
\eq{S\brc{t,\log t}\rightarrow S\brc{t,\log t + y}.}
However, due to the monodromy of the two local solutions $\psi^{\mathrm{out}}_{(\infty)}$ and $\psi^{\mathrm{in}}_{(\infty)}$, the scattering elements remain invariant. In particular, with the shift $\log t \rightarrow \log t + y$ also comes the shift of the logarithm of the local variable $z_L$:
\eq{\log z_L  \rightarrow \log z_L + y,}
which in turn leads to a transformation of the two local solutions
\begin{equation}
\begin{pmatrix}
{\psi}^{\mathrm{out}}_{(\infty)}\\{\psi}^{\mathrm{in}}_{(\infty)}
\end{pmatrix}\rightarrow \mathcal{M}\brc{y} \begin{pmatrix}
\psi^{\mathrm{out}}_{(\infty)}\\
\psi^{\mathrm{in}}_{(\infty)}
\end{pmatrix},
\end{equation}
where $\mathcal{M}\brc{y}$ is a $2\times 2$ monodromy matrix with the following determinant:
\eq{\det \mathcal{M} = \rme^{- y\,t}.}
Now, the full transformation of the scattering elements can be written as 
\eq{S\brc{t,\log t}\rightarrow \frac{\mathcal{M}_{11}\, S\brc{t,\log t+y} + \mathcal{M}_{21}}{\mathcal{M}_{12} \, S\brc{t,\log t+y}+ \mathcal{M}_{22}} =  S\brc{t,\log t},}
where the last equation indicates that the scattering elements should be the same for any branch of the logarithm. This is a powerful statement that allows us to determine the exact form of the scattering elements:
\eq{\frac1{S\brc{t}} = -\frac{t+2\,\mu_{11}}{2\,\mu_{21}} -\frac{\beta}{\mu_{21}} \tanh\brc{\phi+\beta \log t},}
where 
\eq{\mu\equiv
\begin{pmatrix}
\mu_{11} & \mu_{12}\\
\mu_{21} & \mu_{22}
\end{pmatrix}
=\mathcal{M}'\brc{0},\quad \beta = \sqrt{\frac{t^2}{4} -\det \mu},}
and $\phi=\phi\brc{t}$ is the integration constant. As we will show later, $\beta$ is related to the A-period of the corresponding Seiberg-Witten spectral curve, while $\phi\brc{t}$ is related to the B-period.

\section{Regge-Wheeler equation in different local regions}\label{sec:RW}
Our first goal is to find local solutions around the Regge-Wheeler equation's regular and irregular singular points in the small-frequency regime. For that purpose, we need to write down the corresponding differential equations in local variables $z_R$, $z_M$, and $z_L$. We will also perform a wave function transformation to rewrite the Regge-Wheeler equation as a quantum spectral curve, simplifying the solution in the near-horizon region.

Introducing the local variable
\eq{z_R=\frac{2\,M}{r}}
and redefining the radial perturbation $\Phi\brc{r}$ in the Regge-Wheeler equation
\begin{equation}\label{redefinitionschwarzschild}
\Phi(r)=z_R^{-\brc{t+1}/2}  \left(1-z_R\right)^{\frac{t}{2}}\mathrm{e}^{t/(2 z_R) } \psi(z_R),
\end{equation}
we arrive at the differential equation in the near-horizon region:
\begin{equation}
\label{eq:ODE_hor}
\psi_{\text{hor}}''+\frac{z_R (2\, z_R-1)+t}{z_R^2(z_R-1)}\, \psi_{\text{hor}}'+ \frac{\brc{2\ell+1}^2 z_R+\brc{1 - 4\,s^2} z_R^2-2\,t}{4\,z_R^3(z_R-1)}\,\psi_{\text{hor}}=0,
\end{equation}
where $\psi_{\text{hor}}=\psi(z_R)$ is the local wave function.  The boundary condition at the horizon becomes
\eq{\psi_{\text{hor}}= \psi^{\text{in}}_{\text{hor}} \sim 1, \quad z_R\rightarrow 1.}
To set up the perturbative approach described in the previous section, we should be able to write the leading-order solutions of (\ref{eq:ODE_hor}) and their Wronskian in terms of elementary functions. This is indeed possible for integer values of $\ell$ and $s$ such that $\ell\geq s\geq 0$:
\begin{equation}
\begin{aligned}
f^{\text{hor}}_0(z_R)%&=(z_R)^{-\ell-1/2}\,{}_2F_1(-\ell-s,-\ell+s;-2\ell,z_R)
&=\sum_{k=0}^{\ell-s} \frac{(s-\ell)_k\, (-\ell-s)_k}{(-2\ell)_k\, k!} \,z_R^{k-\ell-1/2},\\
g^{\text{hor}}_0(z_R)%&=(z_R)^{\ell+1/2}\, _2F_1(1+\ell-s,1+\ell+s;2+2\ell,z_R)\\
&=\sum_{m=-s}^{\ell-1} a_{s\ell m}\, z_R^{-m-1/2} + \log\brc{1-z_R} \sum_{m=s}^{\ell} b_{s\ell m}\, z_R^{-m-1/2},\\
W^{\text{hor}}_0(z^R)&=\frac{2\,\ell+1}{z_R(1-z_R)},
\end{aligned}
\end{equation}
where the constants $a_{s\ell m}$, $b_{s\ell m}$ are
\begin{equation}
\begin{aligned}
a_{s\ell m}=&\,-b_{s\ell m} \brc{H_{\ell+s}+H_{\ell-s}-H_{m+s} -H_{m-s}},\\
b_{s\ell m}=&\,\frac{\brc{-1}^{\ell+m+1}}{\brc{m+s}!\brc{m-s}!}\, \frac{\brc{2\,\ell+1}!}{\brc{\ell+s}!\brc{\ell-s}!} \frac{\brc{\ell+m}!}{\brc{\ell-m}!}.
\end{aligned}
\end{equation}

The local variable in the near-spatial infinity region is
\eq{z_L=\frac{t}{z_R}.}
The corresponding local wave function $\psi_{(\infty)}=\psi\brc{z_L}$ satisfies the following differential equation:
\begin{equation}
\psi_{(\infty)}''+\frac{z_L+1}{z_L-t}\,\psi_{(\infty)}'+\frac{(2 \ell+1)^2 z_L -2\, z_L^2+t\left(1-4s^2\right)}{4 z_L^2 (t-z_L)}\,\psi_{(\infty)}=0,
\end{equation}
and the boundary condition at the spatial infinity is 
\eq{\psi_{(\infty)} = \psi^{\text{out}}_{(\infty)} \sim \rme^{-z_L} z_L^{-t-1/2}, \quad z_L \rightarrow \infty.}
We will also need the behavior of the incoming wave at the spatial infinity:
\eq{\psi^{\text{in}}_{(\infty)} \sim z_L^{-1/2}, \quad z_L \rightarrow \infty.}
The leading order solutions and their Wronskian in the near-spatial infinity region are
\begin{equation}
\begin{aligned}
f_0^{(\infty)}(z_L)=\mathrm{e}^{-z_L}z_L^{-\ell-1/2}\,p_{\ell}(z_L),&\quad
g_0^{(\infty)}(z_L)=z_L^{-\ell-1/2}\,q_{\ell}(z_L),\\
W_0^{(\infty)}(z_L)\,&= (-1)^{\ell}\,\frac{\mathrm{e}^{-z_L}}{z_L},
\end{aligned}
\end{equation}
where $p_{\ell}(z_L)$ and $q_{\ell}(z_L)$ are the following polynomials of degree $\ell$:
\begin{equation}
\begin{aligned}
p_{\ell}(z_L)=&\,\sum _{m=0}^{\ell} \binom{\ell-\left\lfloor \frac{m+1}{2}\right\rfloor }{\left\lfloor \frac{m}{2}\right\rfloor }\frac{2^{\ell-m}  \left(-2 \left\lfloor \frac{m}{2}\right\rfloor +2 \ell-1\right)\text{!!}}{\left(2 \left\lfloor \frac{m+1}{2}\right\rfloor -1\right)\text{!!}}z_L^m,\\
q_{\ell}(z_L)=&\,\sum _{m=0}^{\ell} \binom{\ell-\left\lfloor \frac{m+1}{2}\right\rfloor }{\left\lfloor \frac{m}{2}\right\rfloor } \frac{2^{\ell-m}  \left(-2 \left\lfloor \frac{m}{2}\right\rfloor +2 \ell-1\right)\text{!!} }{\left(2 \left\lfloor \frac{m+1}{2}\right\rfloor -1\right)\text{!!}}(-z_L)^m.
\end{aligned}
\end{equation}

In the intermediate region, the local variable is
\begin{equation}
z_M=\frac{t^{1/2}}{z_R},
\end{equation}
and the corresponding wave function $\psi_{\text{mid}}=\psi(z_M)$ satisfies the differential equation
\begin{equation}
\psi''_{\text{mid}}+\frac{t^{1/2}\, z_M+1}{z_M-t^{1/2}}\,\psi'_{\text{mid}}+\frac{ (2 \ell+1)^2 z_M+t^{1/2} \left(1-4 s^2-2 z_M^2\right)}{4 z_M^2 \left(t^{1/2}-z_M\right)}\,\psi_{\text{mid}}=0.
\end{equation}
The corresponding leading order solutions and their Wronskian are given by
\begin{equation}
f_0^{\text{mid}}(z_M)=z_M^{\ell+\frac{1}{2}},\quad g_0^{\text{mid}}(z_M)=z_M^{-\ell-\frac{1}{2}},\quad W_0^{\text{mid}}(z_M)=-\frac{2 \ell+1}{z_M}.
\end{equation}

\section{Local wave functions}\label{sec:wavesol}
The multi-polylog approach developed in \cite{Aminov:2023jve} allows us to compute the wave function $\psi_{\text{hor}}$ perturbatively in the parameter $t$ up to any given order (the constraint being the exponential growth of the number of multiple polylogarithm functions entering each order). The particular functions appearing in the near-horizon region are multiple polylogarithms in a single variable 
\eq{\text{Li}_{s_1,\dots,s_n}\brc{z}=\sum_{k_1>k_2>\dots>k_n\geq 1}^{\infty}
\frac{z^{k_1}}{k_1^{s_1}\dots k_{n}^{s_n}},}
which have been extensively studied and have many known properties \cite{goncharov2001multiple,10.1007/978-3-0348-8223-1_1,Lewin'81, Goncharov:1998kja, MINH2000217} . To give a taste of our results, we present a few computed orders of the wave function in the case $\ell=s=0$:
\begin{equation}
\begin{aligned}
\psi_{\text{hor}}^{\text{in}}(z_R)&= z_R^{-1/2}+ \frac{t}{2} \, z_R^{- 1/2}\brc{ \log (z_R)- z_R^{-1}} +t^2\,\psi_2^{\text{hor}}(z_R) +\mathcal{O}\left(t^3\right),\\
\psi_2^{\text{hor}}(z_R)&=-\frac{\text{Li}_2(z_R)}{4\, z_R^{1/2}}+\frac{\log (z_R)\, \text{Li}_1(z_R)}{4\, z_R^{1/2}}+\frac{5\, z_R+4}{24\, z_R^{5/2}}-\frac{(11\, z_R+6) \log (z_R)}{24\, z_R^{3/2}}+\frac{\log ^2(z_R)}{8\, z_R^{1/2}}.
\end{aligned}
\end{equation}
The integration constants $\text{c}_{k}^{R}$ are fixed in each order in $t$ by requiring regularity as $z_R\to 1$ and expressed in terms of multiple zeta values of weight less or equal to $k$. Depending on the quantum numbers, we computed the near-horizon wave function up to orders $t^R_{\text{max}}=13,14,15$.

The wave function in the intermediate region is the simplest one since this region includes two regular singularities at $z_M = 0$ and $z_M = \infty$ in the leading order in $t$. Each order of $\psi_{\text{mid}}$ is a Laurent polynomial in $z_M^{1/2}$ and a polynomial in $\log z_M$. The first few orders of the $t$-expansion in the case $\ell=s=0$ are
\begin{equation}
\psi_{\text{mid}}(z_M)=z_M^{1/2}-\frac{t^{1/2}}{2}\,z_M^{3/2}+\frac{t}{6}\,\,z_M^{1/2}\brc{z_M^2 -3 \log z_M }+\mathcal{O}\left(t^{3/2}\right).
\end{equation}
The integration constants $c_k^M$ are determined by the following continuity condition between the two local wave functions $\psi_{\text{hor}}^{\text{in}}$ and $\psi_{\text{mid}}$:
\eq{\partial_{z_R} \log \brc{\frac{\psi_{\text{hor}}^{\text{in}}}{\psi_{\text{mid}}}}\bigg|_{z_R=\lambda}=0,}
where the gluing point $\lambda$ can be placed anywhere in the interval $\brc{t^{1/2},t^{0}}$, and we consider $t$ real for simplicity (until we start computing the QNMs). Putting $\lambda$ closer to $t^{1/2}$ would increase the maximum order to which the near-horizon wave function needs to be computed. Thus, we choose $\lambda=t^{1/6}$, which is far enough from $t^{1/2}$ and not too close to $t^{0}$. With this choice of $\lambda$, the maximum required order in $t$ for the intermediate wave function is at least $t^M_{\text{max}}\sim 5\,t^R_{\text{max}} /2$.

Since radial infinity is an irregular singular point of the differential equation, we must modify the multi-polylog approach to find the local wave function in the near-spatial infinity region.
Instead of multiple polylogarithms, new special functions appear in the perturbative expansion - multiple polyexponential integrals $\mathrm{ELi}_{s_1,\dots,s_n}(z)$. These functions can be defined as iterated integrals of exponential integral $\mathrm{Ei}(z)$ and are much less studied (for the study of related functions see \cite{boyadzhiev2007polyexponentials,kim2020degenerate,Kim2019ANO,KIM2020124017,komatsu1,komatsu2,lacpao2019hurwitz}). As with multiple polylogarithms, the level of $\mathrm{ELi}_{s_1,\dots,s_n}(z)$ is $n$ and the weight is $s_1+\dots+s_n$.
The first special function of this kind is the exponential integral itself:
\begin{equation}\label{Eiandseries}
\mathrm{ELi}_{1}(z)\equiv\mathrm{Ei}(z)=\gamma+\log(-z)+\sum_{k=1}^{\infty}\frac{z^k}{k!\,k},\quad |\mathrm{Arg}(-z)|<\pi.
\end{equation}
The functions with higher weight are defined recursively as
\begin{equation}\label{recurrenceELi}
\begin{aligned}
s_1=1\colon\quad \mathrm{ELi}_{1,s_2,\dots,s_n}(z)&=-\int_{-\infty}^z\frac{\mathrm{e}^t}{t}\,\mathrm{ELi}_{s_2,\dots,s_n}(-t)\,\mathrm{d}t,\\
s_1>1\colon\quad \mathrm{ELi}_{s_1,s_2,\dots,s_n}(z)&=\int_{-\infty}^z\frac{1}{t}\,\mathrm{ELi}_{s_1-1,s_2,\dots,s_n}(t)\,\mathrm{d}t.
\end{aligned}
\end{equation}
All the necessary properties of these functions can be found in a separate work \cite{Aminov:2024aan}.

The solution in the near-spatial infinity region can be written as a linear combination of the purely outgoing wave $\psi_{(\infty)}^{\text{out}}$ and the purely incoming wave $\psi_{(\infty)}^{\text{in}}$:
\eq{\psi_{(\infty)} = \mathcal{A}\,\psi^{\mathrm{out}}_{(\infty)}+\mathcal{B}\,\psi^{\mathrm{in}}_{(\infty)}.}
The first few orders of this solution in the case of $\ell=s=0$ are
\begin{align}\label{outgoingwave}
\psi_{(\infty)}^{\text{out}}(z_L)&=\mathrm{e}^{-z_L}\,z_L^{-1/2}+t\,\psi_{1,\text{out}}^{(\infty)}(z_L)+t^2\,\psi_{2,\text{out}}^{(\infty)}(z_L)+\mathcal{O}\left(t^3\right),\\
\psi_{1,\text{out}}^{(\infty)}(z_L)&=\frac{1}{2}\,z_L^{-1/2}\, \text{ELi}_1(-z_L)+\frac{1}{2}\, \mathrm{e}^{-z_L} \,z_L^{-3/2}-\mathrm{e}^{-z_L}\, z_L^{-1/2} \, \log z_L,\nonumber\\
\psi_{2,\text{out}}^{(\infty)}(z_L)&=\frac{1}{4 \,z_L^{1/2}}\left[\,\mathrm{e}^{-z_L}\,\text{ELi}_{1,1}(z_L)+2\,\text{ELi}_2(-z_L)\right]-\frac{17\,z_L-6}{24\,z_L^{3/2}}\,\text{ELi}_1(-z_L) +\nonumber\\
&\ \ +\frac{\mathrm{e}^{-z_L}}{2\,z_L^{1/2}}\,\log(z_L)\left[\log(z_L)-\mathrm{e}^{z_L}\,\mathrm{ELi}_1(-z_L)-z_L^{-1}\right]+ \mathrm{e}^{-z_L} \, \frac{13\, z_L+8}{24\,z_L^{5/2}} ,\nonumber\\
\label{incomingwave}
\psi_{(\infty)}^{\text{in}}(z_L)&=z_L^{-1/2}+t\,\psi_{1,\text{in}}^{(\infty)}(z_L)+t^2\,\psi_{2,\text{in}}^{(\infty)}(z_L)+\mathcal{O}\left(t^3\right),\\
\psi_{1,\text{in}}^{(\infty)}(z_L)&=\frac{1}{2}\, z_L^{-3/2}-\frac{1}{2}\,\mathrm{e}^{-z_L}\, z_L^{-1/2}\,\text{ELi}_1(z_L),\nonumber\\
\psi_{2,\text{in}}^{(\infty)}(z_L)&=\frac{\text{ELi}_{1,1}(-z_L)}{4 \,z_L^{1/2}}+\frac{\mathrm{e}^{-z_L}}{24\,z_L^{1/2}}\left[\mathrm{e}^{z_L}\frac{11\,z_L+8}{z_L^2}-\frac{17\,z_L+6}{z_L} \,\text{ELi}_1(z_L) +12\,\text{ELi}_2(z_L)\right].\nonumber
\end{align}
To fix the integration constants $c_k^L$ and determine the coefficients $\mathcal{A}$ and $\mathcal{B}$, we impose the continuity condition:
\eqlb{eq:Cont_ML}{\partial_{z_L} \log \brc{\frac{\psi_{(\infty)}}{\psi_{\text{mid}}}}\bigg|_{z_L=\lambda}=0,}
where $\lambda = t^{1/6}$. In the above equation, one needs to know how polyexponential integrals (\ref{recurrenceELi}) behave as $z\rightarrow 0$. As we show in \cite{Aminov:2024aan}, $\mathrm{ELi}_{s_1,\dots,s_n}(z)$ admit relations similar to (\ref{Eiandseries}) with another set of functions we call \emph{undressed multiple polyexponential functions} $el_{s_1,\dots,s_n}(z)$ defined as
\begin{equation}
el_{s_1,s_2,\dots,s_n}(z)=\sum_{k_1>k_2>\dots>k_n\ge 1}\frac{z^{k_1}}{k_1!\,k_1^{s_1}\,k_2^{s_2}\dots k_n^{s_n}}.
\end{equation}
As a result of (\ref{eq:Cont_ML}), we obtain $\mathcal{A}$ and $\mathcal{B}$ in the form of series expansions in the parameter $t$, which also depend on the logarithm of $t$. Computing $\mathcal{B}/\mathcal{A}$ gives us the perturbative expansion of the inverse scattering elements $S_{\ell,s}^{-1}$. The first few orders in the case of $\ell=s=0$ are
\begin{equation}\label{eq:inverseGreen_l=s=0}
\begin{aligned}    S_{0,0}^{-1} =&\,-1+\frac{1}{2} (1 -\imath\,\pi -2 \gamma)\brc{1+\gamma\,t+\frac{\gamma^2}{2}\,t^2} t+\frac{1}{24}\brc{9-5\,\imath\, \pi +3\,\pi^2 + 12\,\gamma^2} \brc{1+\gamma\,t}t^2 \\
&\,+ \frac{1}{144} \left(-15+78\,\imath\, \pi +35\,\pi^2+ 3\,\imath\, \pi^3 -36 \log t-  12\, \zeta(3)  -36\, \gamma -24\, \gamma^3 \right) t^3 +\mathcal{O}\brc{t^4},
\end{aligned}
\end{equation}
where $\gamma$ is the Euler–Mascheroni constant. The maximum order to which we were able to compute the inverse scattering elements is $14$ for $\ell=s=0$ and $13$ for $\ell=s=1,2$. At the same time, the near-spatial infinity wave function was computed up to order $t^L_{\text{max}}=16$ for $\ell=s=0$ and up to $t^L_{\text{max}}=15$ for higher values of $\ell$ and $s$. The first few orders of the inverse scattering elements in the cases $\ell=s=1,2$ are
\begin{equation}\label{eq:inverseGreen_l=s=1}
\begin{aligned}    
S_{1,1}^{-1} =&\,1+\frac{1}{4} \brc{ 2\, \imath\, \pi-5+4\, \gamma} \brc{1+\gamma\,t+\frac{\gamma^2}{2}\,t^2} t+ \frac{1}{32} \left(25-16 \gamma^2-\frac{206\,\imath\, \pi }{15}-4\, \pi ^2\right) \brc{1+\gamma\,t} t^2\\
& +\frac{1}{96} \left(-33+16\, \gamma^3+14\, \imath\, \pi +\frac{37\,\pi ^2}{15}-2\, \imath\, \pi ^3+8\,\zeta (3)\right) t^3 +\mathcal{O}\brc{t^4},
\end{aligned}
\end{equation}
\begin{equation}\label{eq:inverseGreen_l=s=2}
\begin{aligned}    S_{2,2}^{-1}=&\,-1+\frac{1}{6} \brc{10-6 \,\gamma -3\, \imath\, \pi} \brc{1+\gamma\,t+\frac{\gamma^2}{2}\,t^2} t+ \frac{1}{72} \left(-100+36\, \gamma ^2+\frac{1779\, \imath \,\pi }{35}+9\, \pi^2\right) \brc{1+\gamma\,t} t^2\\
& +\frac{1}{96} \left(73-16\, \gamma^3-\frac{324\, \imath \, \pi }{7}-\frac{1244 \,\pi^2}{105}+2\, \imath\, \pi^3 -8\,\zeta(3)\right) t^3 +\mathcal{O}\brc{t^4}.
\end{aligned}
\end{equation}

\section{Monodromy in the near-spatial infinity region}\label{sec:monodromy}

Having computed the perturbative expansion of the near-spatial infinity wave function, we can study the monodromy of the two solutions $\psi^{\mathrm{out}}_{(\infty)}$  and $\psi^{\mathrm{in}}_{(\infty)}$. In the near-spatial infinity region, when $t\rightarrow 0$, the two regular singularities at $r=0$ ($z_L=0$) and $r=2\,M$ ($z_L=t$) are very close. A small oriented loop around $z_L=0$ incircling both singularities is equivalent to a loop around spatial infinity with opposite orientation. Thus, the only monodromy matrix available to us in the near-spatial infinity region is $\mathcal{M}\equiv \mathcal{M}_{(\infty)}$. Changing the branch of the logarithm by
\eq{\log z_L  \rightarrow \log z_L + y,}
induces the following monodromy transformation of local solutions:
\begin{equation}
\label{eq:Mon_trans}
\begin{pmatrix}
{\psi}^{\mathrm{out}}_{(\infty)}\\{\psi}^{\mathrm{in}}_{(\infty)}
\end{pmatrix}\rightarrow \mathcal{M}\brc{y} \begin{pmatrix}
\psi^{\mathrm{out}}_{(\infty)}\\
\psi^{\mathrm{in}}_{(\infty)}
\end{pmatrix},
\end{equation}
where the monodromy matrix is a $2\times 2$ matrix:
\eq{\mathcal{M}(y)=\begin{pmatrix}
\mathcal{M}_{11}  & \mathcal{M}_{12}\\
\mathcal{M}_{21} & \mathcal{M}_{22}
\end{pmatrix}.}
Here, we will treat the parameter $y$ as a continuous variable, not necessarily equal to $2\,\pi \imath\, n$, $n\in \mathbb{Z}$. Thus, $\mathcal{M}$ is an element of a one-parameter Lie group with an identity element at $y=0$. The corresponding Lie algebra generator is 
\eq{\mu= \mathcal{M}'(0).}
The determinant of $\mathcal{M}$ is
\eq{\det \mathcal{M} = \rme^{- y\,t},}
which results in 
\eq{\Tr \mu= -t.}
Similar monodromy matrices were studied earlier using other methods in the context of the confluent Heun equation or Painlevé V equation  \cite{Castro:2013kea, Castro:2013lba, CarneirodaCunha:2015hzd, lisovyy2018irregular, CarneirodaCunha:2019tia}. An important question we do not answer here is a matching between our results and those in the literature. One of the simpler differences is an additional factor of $\rme^{- y\,t/2}$ in front of the monodromy matrix as compared to the one in \cite{lisovyy2018irregular}, which is due to the wave function transformation.

Each element of the monodromy matrix is obtained as a series expansion in $t$, which we computed up to order $t^L_{\text{max}}$. These perturbative expansions allow us to determine the exact dependence of $\mathcal{M}$ on the parameter $y$:
\begin{equation}
\label{eq:Mon_elem}
\begin{aligned}
\mathcal{M}_{11}&=\rme^{-y\,t/2} \left(\cosh (y\,\beta)+\frac{\left(\mu _{11}-\mu_{22}\right)}{2 \beta } \sinh (y\, \beta)\right),\quad 
\mathcal{M}_{12}=\rme^{-y\,t/2} \, \frac{\mu_{12}}{\beta} \sinh (y\, \beta),\\
\mathcal{M}_{22}&=\rme^{-y\,t/2} \left(\cosh (y\,\beta)-\frac{\left(\mu _{11}-\mu_{22}\right)}{2 \beta } \sinh (y\, \beta)\right),\quad \mathcal{M}_{21}=\rme^{-y\,t/2} \, \frac{\mu_{21}}{\beta} \sinh (y\, \beta),
\end{aligned}
\end{equation}
where $\mu_{ij}$ are the elements of the generator $\mu$, and $\beta$ is related to the determinant of  $\mu$:
\eq{\beta = \sqrt{\frac{t^2}{4} -\det \mu}.}
In (\ref{eq:Mon_elem}), the parameters that depend on quantum numbers $\ell$ and $s$ are $\mu_{ij}$ and $\beta\equiv \beta_{\ell,s}$.

The perturbative expansions of the two local solutions allow us to determine the elements of $\mu$ up to the same order $t^L_{\text{max}}$. For example, the first few orders in the $t$-expansion of $\mu_{11}$ in the case $\ell=s=0$ are
\eq{\mu_{11}=-t+\frac{\imath\, \pi}{4}\,   \,t^2+\frac{\pi ^2}{12}\,t^3+\mathcal{O}\left(t^4\right).}
%\begin{equation}
%\begin{aligned}
%\mu_{11}&=-t+\frac{1}{4}\, \imath\, \pi  \,t^2+\frac{\pi ^2}{12}\,t^3+\mathcal{O}\left(t^4\right),\\
%\mu_{12}&=\frac{t}{2}+\frac{1}{24}\, (12\, \gamma -17)\, t^2+\frac{1}{48} \left(20-34\, \gamma +12\, \gamma ^2-\pi ^2\right)\,t^3+\mathcal{O}\left(t^4\right),\\
%\mu_{21}&=-\frac{t}{2}+\frac{1}{24} (-17+12\, \gamma +12\, \imath\, \pi )\,t^2+\frac{1}{48} \left(-20+34\, \gamma -12\, \gamma ^2+34\, \imath\, \pi -24\, \imath\, \gamma\,  \pi +13\, \pi ^2\right)\,t^3+\mathcal{O}\left(t^4\right),\\
%\mu_{22}&=-\frac{1}{4}\, \imath\, \pi\,  t^2-\frac{\pi ^2 }{12}\, t^3+\mathcal{O}\left(t^4\right).
%\end{aligned}
%\end{equation}
The corresponding Taylor expansion for $\beta_{\ell,s}$ is 
\begin{equation}
\beta_{0,0}=\frac{7}{24}\, t^2-\frac{9449}{120\,960}\,t^4+\frac{102\,270\,817 }{2\,133\,734\,400}\,t^6+\mathcal{O}\left(t^{8}\right).
\end{equation}

It is possible to find the exact expressions for $\mu_{11}$ and $\mu_{12}$ in terms of $\beta_{\ell,s}$ and a new function $\xi_{\ell, s}=\xi_{\ell, s}(t)$:
\begin{equation}\label{mu11}
\mu_{11}(t)=-\frac{t}{2}-\imath\,\beta\,\frac{\mathrm{e}^{-\imath\,\pi\,t}-\cos(2\,\pi\,\beta)}{\sin(2\,\pi\,\beta)},
\end{equation}
\begin{equation}
\mu_{12}(t)=\frac{2\,\pi\,\beta}{\sin\brc{2\,\pi\,\beta} } \frac{\brc{\frac{t}{2}-\beta} \xi_{\ell,s}}{\Gamma\brc{1+\frac{t}{2}+\beta} \Gamma\brc{1+\frac{t}{2}-\beta}},
\end{equation}
where $\xi_{\ell, s}$ satisfies
\begin{equation}
\xi_{\ell,s}(t)\,\xi_{\ell,s}(-t)=1.
\end{equation}
In the case $\ell=s=0$, $\xi_{\ell, s}$ admits the following Taylor expansion:
\begin{equation}
\xi_{0,0}(t)=1-\frac{5}{6}\,t+\frac{25}{72}\,t^2-\frac{23}{315}\,t^3+\frac{41}{72\,576}\,t^4+\mathcal{O}\left(t^5\right).
\end{equation}
The remaining elements $\mu_{ij}$ can be obtained using the trace and the determinant of $\mu$:
\eq{\mu_{11}+\mu_{22}=-t,\quad
\mu_{12} \, \mu_{21} = \mu_{11}\,\mu_{22} +\beta^2-\frac{t^2}{4}.}

Following \cite{lisovyy2018irregular}, we look at the eigenvalues of the Monodromy matrix $\mathcal{M}$ rescaled by $\rme^{\,y\,t/2}$:
\eq{\rme^{\,y\,t/2} \mathcal{M}= U
\begin{pmatrix}
\rme^{-\beta\,y}  & 0\\
0 & \rme^{\,\beta\,y}
\end{pmatrix}
U^{-1},}
which tells us that $\beta$ is related to the flat modulus $a$ of the corresponding Seiberg-Witten curve. However, $\beta$ is obtained by resumming instanton corrections using the dictionary between the gauge theory quantities and the black hole parameters. To be more precise, $a$ can be computed by inverting the Matone relation \cite{Matone:1995rx} perturbatively in the instanton parameter $\Lambda\equiv t$. The resulting instanton expansion depends on both $\Lambda$ and the frequency $\omega$ (also proportional to $t$ via (\ref{eq:SW_BH_dict})). In the $\ell=s=0$ case, the following expansion for $a$ can be derived:
\eqlb{eq:a_Lambda_exp}{a= \frac{1}{2} \sqrt{8\, \omega ^2-1} +\frac{\imath \,  \omega \,\Lambda }{4 \sqrt{8\, \omega^2-1}} +\frac{ \left(272\, \omega ^4+70\, \omega ^2-11\right)\Lambda ^2}{64 \left(8\,\omega^2-1\right)^{3/2} \left(8\, \omega^2+3\right)} +\mathcal{O}\brc{\Lambda^3},}
where we fixed $M=1/2$ for simplicity.
When rewriting this expansion using the single parameter $t$, \emph{infinitely} many orders in $\Lambda$ contribute to the same order in $t$. For example, one can check the following:
\eq{\imath\, a=-\ell-\frac{1}{2} + \mathcal{O}\brc{t^2}.}
This leads us to the relation between $a$ and $\beta_{\ell,s}$:
\eqlb{eq:a_beta_rel}{\imath \,a=- \ell- \frac{1}{2} - \beta_{\ell,s}.}
To check this claim, we can approximately compute coefficients in front of $t^2$ and $t^4$ for $\ell=s=0$ using $15$ instanton orders in (\ref{eq:a_Lambda_exp}):
\eq{\imath\, a\simeq -\frac{1}{2} -0.29170\, t^2+ 0.07816\, t^4+\mathcal{O}\brc{t^6},}
which are indeed close to $-7/24 \simeq -0.29167$ and $9449/120\,960 \simeq 0.07812$.
The resummation of infinitely many instantons comes from the poles of the Nekrasov partition function in $a$, and it starts at order $t^{2\ell+2}$ for higher values of $\ell$. This means, for example, that for $\ell=1$, the first two instantons provide an exact expression for $a$ up to order $t^2$. Indeed, we have for $\ell=s=1$
\eq{\imath \, a= -\frac{3}{2} -\frac{47}{240}\, t^2 +\mathcal{O}\brc{t^4}}
and
\eq{\beta_{1,1}= \frac{47}{240}\, t^2 -\frac{43\,908\,007}{1\,137\,024\,000} \, t^4 +\mathcal{O}\brc{t^6}.}
In Appendix \ref{appendix:beta}, we use instanton expansion to derive generic formulas for coefficients in front of $t^{2\,k}$, $k=1,2,3$ in the expansion of $\beta_{\ell,s}$, which are valid for $\ell\geq k$.

\section{Scattering elements}\label{sec:amplitudes}

Applying the monodromy transformation (\ref{eq:Mon_trans}) to the full solution in the near-spatial infinity region
\eq{
\psi_{(\infty)} = \mathcal{A}\,\psi^{\mathrm{out}}_{(\infty)}+\mathcal{B}\,\psi^{\mathrm{in}}_{(\infty)},}
allows us to understand how the scattering elements $S=\mathcal{A}/\mathcal{B}$ behave under the shift of the logarithm $\log t \rightarrow \log t + y$:
\eq{S\brc{t,\log t}\rightarrow \frac{\mathcal{M}_{11}\, S\brc{t,\log t+y} + \mathcal{M}_{21}}{\mathcal{M}_{12} \, S\brc{t,\log t+y}+ \mathcal{M}_{22}}.}
Since changing the branch of the logarithm should not affect any physical quantities, the following functional relation should hold: 
\eqlb{eq:Green_trans1}{ \frac{\mathcal{M}_{11}\, S\brc{t,\log t+y} + \mathcal{M}_{21}}{\mathcal{M}_{12} \, S\brc{t,\log t+y}+ \mathcal{M}_{22}} =  S\brc{t,\log t},}
which we can verify in the small $t$ expansion.
It follows that the $y$-dependence is only apparent:
\eq{ \frac{\mathrm{d}}{\mathrm{d}y}\left[\frac{\mathcal{M}_{11}\, S\brc{t,\log t+y} + \mathcal{M}_{21}}{\mathcal{M}_{12} \, S\brc{t,\log t+y}+ \mathcal{M}_{22}}\right]\bigg|_{y=0} = 0.}
At $y=0$, the monodromy matrix is an identity matrix
\begin{equation}
\mathcal{M}_{11}|_{y=0}=\mathcal{M}_{22}|_{y=0}=1,\quad \mathcal{M}_{12}|_{y=0}=\mathcal{M}_{21}|_{y=0}=0,
\end{equation}
which leads to the following first-order differential equation on the inverse scattering elements:
\begin{equation}
\label{eq:G_diff}
\frac{\partial}{\partial\log t}\,\frac{1}{S\brc{t,\log t}}= \frac{\mu_{21}}{S\brc{t,\log t}^2}+ \frac{\mu_{11}-\mu_{22}}{S\brc{t,\log t}}-\mu_{12}.
\end{equation}
The exact dependence of the scattering elements on $\log t$ can thus be derived:
\eqlb{exactGreen}{\boxed{\frac1{S\brc{t,\log t}} = -\frac{t+2\,\mu_{11}}{2\,\mu_{21}} -\frac{\beta}{\mu_{21}} \tanh\brc{\phi+\beta \log t},}}
where the integration constant is a function of $t$: $\phi=\phi(t)$. Computing $\phi(t)$ up to a certain order in $t$, we verify that it is related to the B-period of the corresponding Seiberg-Witten curve with some additional contributions. In terms of $\beta$, we have:
\begin{equation}
\label{eq:phi_exp_beta}
\begin{aligned}
\phi(t) =  \imath \,\pi\,\beta+ \brc{\ell+\frac{1}{2}} \log t + \log \frac{\Gamma\brc{-2\,\ell -2\,\beta}}{\Gamma\brc{2\,\ell +2\,\beta+2}} - \frac{1}{2} \sum_{j=1}^{3} \log \frac{\Gamma\brc{m_j-\ell-\beta}}{\Gamma\brc{m_j+\ell+\beta+1}}& \\
-\frac{1}{2} \log \frac{\sin\bsq{\pi\brc{\beta-t/2}}}{\sin\bsq{\pi\brc{\beta+t/2}}}+ \sum_{k\geq 0} \varphi_{\ell,s,k} \,t^{2\,k} &,
\end{aligned}
\end{equation}
where the hypermultiplet masses $m_j$ are
\eq{m_1=\frac{t}{2}+s,\quad m_2=\frac{t}{2}-s, \quad m_3=\frac{t}{2},}
and the coefficients $\varphi_k$ can be computed perturbatively. The first few coefficients in the case of $\ell=s=0$ are
\eqlb{eq:varphi_00}{\sum_{k\geq 0} \varphi_{0,0,k} \,t^{2\,k} = \log\brc{ \frac{7}{9}} -\frac{8587}{70\,560}\, t^2 +\frac{59\,423\,233}{995\,742\,720} \, t^4+\mathcal{O}\brc{t^6}.}
One might notice that $\phi\brc{t}$ depends on $\brc{\ell+1/2}\log t$, which would seemingly violate the differential equation (\ref{eq:G_diff}). However, $1/2\,\log t $ is canceled by the expansion of the log-Gamma functions, and the contribution of $\ell \log t$ can be rewritten as
\eqlb{eq:phi_l_log}{\tanh\brc{\ell \log t + x} =\frac{t^{2\,\ell} \,\rme^{2\,x}-1}{t^{2\,\ell}\, \rme^{2\,x}+1},}
which makes it invisible for the partial derivative with respect to $\log t$ in the small $t$ expansion. Another consequence of (\ref{eq:phi_l_log}) is that the contribution of $\log\brc{t}$ to the expansion of the inverse scattering elements is delayed to order $t^{2\,\ell+3}$, in agreement with (\ref{eq:inverseGreen_l=s=0}).

Let us comment on the relation with the Seiberg-Witten B-period in more detail. We can see that $\phi\brc{t}$ contains the perturbative part of the B-period, which is given by the $a$-derivative of the Nekrasov-Shatashvili (NS) free energy with $N_f=3$ and $\hbar=1$:
\eq{\frac{\partial\, \mathcal{F}^{(\text{NS})}_{\text{pert}}}{\partial\, a} = -2\,a \log t - 2\,\imath \log \frac{\Gamma\brc{1 + 2\,\imath\, a}}{\Gamma\brc{1 - 2\,\imath\, a}}+ \imath\sum_{j=1}^{3} \log \frac{\Gamma\brc{m_j + \imath\, a+1/2}}{\Gamma\brc{m_j - \imath\, a+1/2}} .}
Taking into account relation (\ref{eq:a_beta_rel}) between $a$ and $\beta$, we can write for $\phi\brc{t}$:
\eqlb{eq:phi+betalog}{\phi\brc{t}+\beta \log t= \imath \,\pi\,\beta +\frac{\imath}{2} \frac{\partial\, \mathcal{F}^{(\text{NS})}_{\text{pert}}}{\partial\, a} -\frac{1}{2} \log \frac{\sin\bsq{\pi\brc{\beta-t/2}}}{\sin\bsq{\pi\brc{\beta+t/2}}} + \sum_{k\geq 0} \varphi_{\ell,s,k} \,t^{2\,k} .}
This leads us to think that the sum
\eq{\varphi\equiv\varphi_{\ell,s}=\sum_{k\geq 0} \varphi_{\ell,s,k} \,t^{2\,k}}
is related to the instanton part of the NS free energy. To confirm this relation, we compute the $a$-derivative of the instanton expansion and substitute (\ref{eq:a_Lambda_exp}) to get for $\ell=s=0$:
\eq{\frac{\partial\, \mathcal{F}^{(\text{NS})}_{\text{inst}}}{\partial\, a}=\frac{\sqrt{8\, \omega ^2-1}}{8\, \imath\,\omega }\,\Lambda +\frac{ \left(-1920\, \omega ^6-912\, \omega ^4+40\, \omega ^2+45\right) \Lambda ^2}{256\, \omega ^2 \sqrt{8\,\omega ^2-1} \left(8\, \omega ^2+3\right)^2}+\mathcal{O}\brc{\Lambda^3}.}
Rewriting this as an expansion in $t$ will result in resumming infinitely many orders in $\Lambda$. Using $15$ instanton corrections, we get approximately:
\eq{\frac{\imath}{2} \,\frac{\partial\, \mathcal{F}^{(\text{NS})}_{\text{inst}}}{\partial\, a} \simeq -0.25095 -0.12213\, t^2 +0.05985\, t^4+\mathcal{O}\brc{t^6},}
where only even powers of $t$ are present consistently with (\ref{eq:varphi_00}).
All the numeric coefficients in the above instanton resummation also agree with (\ref{eq:varphi_00}), where
\eq{\varphi_{0,0,0} \simeq -0.25131, \quad  \varphi_{0,0,1} \simeq -0.12170,\quad 
\varphi_{0,0,2} \simeq 0.05968.}
In this case, the resummation of instantons starts at order $t^{2\ell}$, which is due to the $a$-derivative that increases the order of the poles in each instanton contribution. Thus, for $\ell=2$, the first two instantons should be enough to match the coefficient $\varphi_{2,s,1}$ exactly. Taking $\ell=s=2$, we get
\eq{\frac{\imath}{2} \,\frac{\partial\, \mathcal{F}^{(\text{NS})}_{\text{inst}}}{\partial\, a}= \frac{125}{1568}\, t^2+\mathcal{O}\brc{t^4},}
which agrees with the expansion for $\varphi_{2,2}$:
\eq{\varphi_{2,2}= \frac{125}{1568}\, t^2-\frac{53\,950\,959\,337}{2\,280\,051\,680\,256} t^4+\mathcal{O}\brc{t^6}.}

Plugging the relation between $\varphi_{\ell,s}$ and the instanton expansion back into (\ref{eq:phi+betalog}), gives
\eq{\phi\brc{t}+\beta \log t= \imath \,\pi\,\beta +\frac{\imath}{2} \frac{\partial\, \mathcal{F}^{(\text{NS})}_{\text{pert}}}{\partial\, a} +\frac{\imath}{2} \,\frac{\partial\, \mathcal{F}^{(\text{NS})}_{\text{inst}}}{\partial\, a} -\frac{1}{2} \log \frac{\sin\bsq{\pi\brc{\beta-t/2}}}{\sin\bsq{\pi\brc{\beta+t/2}}} ,}
which provides an alternative description for the scattering elements in terms of the NS free energy and the flat modulus $a$. 

\section{QNM frequencies}\label{sec:QNM}

The poles of the scattering elements obtained in the previous section determine the quasinormal mode frequencies $\omega_n\equiv \omega_{n,\ell,s}$, provided the relation between $t$ and $\omega$ given in (\ref{eq:SW_BH_dict}). The corresponding condition for the inverse scattering elements is
\eqlb{eq:QNM_quant}{\frac{1}{S\brc{-4\,\imath\, M \omega_n}}=0,}
which we call the \emph{quantization condition} for short. The latter can be rewritten with the help of \eqref{exactGreen} as
\begin{equation}
\frac{\mathrm{e}^{2\,\phi+2\,\beta\,\log(t)}-1}{\mathrm{e}^{2\,\phi+2\,\beta\,\log(t)}+1}=-\frac{t+2\,\mu_{11}}{2\,\beta}.
\end{equation}
Substituting the element of the monodromy generator \eqref{mu11}, we get
\begin{equation}
\label{eq:GI_zero_sim}
\mathrm{e}^{2\, \phi+2\,\beta\,\log(t)}=\frac{e^{\imath\, \pi\left(t+2\, \beta\right)}-1}{1-e^{\imath\, \pi \left(t-2\, \beta \right)}}=\mathrm{e}^{2\,\imath\, \pi\, \beta} \,\frac{ \sin \left[\pi  \left(\beta +\frac{t}{2}\right)\right]}{\sin \left[\pi  \left(\beta -\frac{t}{2}\right)\right]}.
\end{equation}
Now, we can use the expression (\ref{eq:phi_exp_beta}) for $\phi\brc{t}$ in terms of $\beta$, where $\imath\,\pi\,\beta$ and the logarithm of sines conveniently cancel with the \textit{rhs} of (\ref{eq:GI_zero_sim}):
\begin{equation}
\label{eq:quant_Gamma_1}
\boxed{
\rme^{2\,\varphi} t^{2\,\ell+2\,\beta+1} \,\frac{\Gamma\brc{-2\,\ell -2\,\beta}^2}{\Gamma\brc{2\,\ell +2\,\beta+2}^2}\prod_{j=1}^3\frac{\Gamma\brc{m_j+\ell+\beta+1}}{\Gamma\brc{m_j-\ell-\beta}}=1.}
\end{equation}
By taking the logarithm, we can also rewrite (\ref{eq:quant_Gamma_1}) in a form equivalent to the quantization of the Seiberg-Witten B-period:
\begin{equation}
\label{eq:quant_SW}
\boxed{
\frac{\partial\, \mathcal{F}^{(\text{NS})}_{\text{pert}}}{\partial\, a} + \frac{\partial\, \mathcal{F}^{(\text{NS})}_{\text{inst}}}{\partial\, a} =2\,\pi\brc{n+1},\quad n\in\mathbb{Z}_{\ge 0},
}
\end{equation}
where $a$ is related to $\beta$ via (\ref{eq:a_beta_rel}). 

Quantization condition (\ref{eq:quant_Gamma_1}) requires a simple input in the form of two Taylor expansions in $t$, where only even powers of $t$ are present:
\eqlb{eq:beta_phi_exp}{\beta_{\ell,s}=\sum_{k\geq 1} \beta_{\ell,s,k} \, t^{2k},\quad 
\varphi_{\ell,s} = \sum_{k\geq 0} \varphi_{\ell,s,k} \,t^{2\,k}.}
For given quantum numbers $\ell$ and $s$, the coefficients $\beta_{\ell,s,k}$ and $\varphi_{\ell,s,k}$ are rational numbers, except for $\varphi_{0,0,0} = \log\brc{7/9}$. Moreover, it is possible to derive generic formulas for $\beta_{\ell,s,k}$, $\ell\geq k$ and $\varphi_{\ell,s,k}$, $\ell>k$ from the instanton expansion  (see Appendix \ref{appendix:beta}). Since $\beta$ is entirely determined by the monodromy in the near-spatial infinity region, the coefficients $\beta_{\ell,s,k}$ can be computed solely from the perturbative expansions of $\psi^{\mathrm{out}}_{(\infty)}$  and $\psi^{\mathrm{in}}_{(\infty)}$. The second Taylor expansion $\varphi_{\ell,s}$ is related to the Seiberg-Witten B-period and thus also requires the knowledge of the near-horizon wave function $\psi^{\text{in}}_{\text{hor}}$.

The radius of convergence $r$ of the $t$-expansion of $\beta_{\ell,s}$ seems to depend on the angular quantum number $\ell$: $r\equiv r_{\ell}$. For $\ell =0$, we have approximately $\abs{t}<r_{0}\sim 1$. This radius might be related to the poles in the instanton expansion for $a$ (\ref{eq:a_Lambda_exp}). The first pole appears at order $\Lambda^2$ and is of the form $8\,\omega^2 +3$, which is equivalent to $3- 2\,t^2$, provided $M=1/2$. If this is true, then $r_0$ is indeed close to one: $r_0=\sqrt{3/2}$. However, we don't know if the poles in the instanton expansion are meaningful for the black hole perturbation theory or if they are remnants of the inversion of the Matone relation. Acknowledging its speculative nature, one can continue the above argument for higher values of  $\ell$. The poles in each order of the instanton expansion are of the form $2\,\imath\,a\pm k$, $k\in \mathbb{N}$. We have for $a$ in the leading order:
\eq{a^2=2\,\omega^2 -\brc{\ell+\frac{1}{2}}^2 +\mathcal{O}\brc{\Lambda},}
which corresponds to the poles in $t$ at
\eq{t^2=\frac{k^2-\brc{2\,\ell+1}^2}{2}.}
As we mentioned earlier, the pole at $t=0$ leads to the resummation of the instanton expansion into (\ref{eq:beta_phi_exp}). The next pole closest to $t=0$ is given by $k=2\,\ell$ with $\ell>0$. Thus, we get the following estimation for the radius of convergence of $\beta_{\ell,s}$ in (\ref{eq:beta_phi_exp}):
\eqlb{eq:r_conv_ell}{\ell>0:\quad r_{\ell}=\sqrt{\frac{4\,\ell+1}{2}}.}
By taking the square root of the ratios $\abs{\beta_{\ell,s,k}/\beta_{\ell,s,k+1}}$, we can also estimate the radius of absolute convergence $r^{\text{abs}}_{\ell}$ for $\ell=0,1,2$:
\eq{r^{\text{abs}}_0\simeq 1, \quad 
r^{\text{abs}}_1\simeq 1.3,\quad  r^{\text{abs}}_2\simeq 1.6.}
For comparison, we have
\eq{r_0\simeq 1.2,\quad r_1\simeq 1.6,\quad
r_2\simeq 2.1.}

The above $r_{\ell}$ values allow us to compute the fundamental frequencies $\omega_{0,\ell,s}$ via (\ref{eq:quant_Gamma_1}) or (\ref{eq:quant_SW}). Using the $t$-expansions of $\beta_{\ell,s}$ and $\varphi_{\ell,s}$ from Appendix \ref{appendix:beta}, we get:
\eq{\omega_{0,0,0}= \pm {\bf 0.22091}1 - {\bf 0.20979}2 \imath ,\quad 
\omega_{0,1,1}= \pm {\bf 0.497}254 - {\bf 0.185}939 \imath ,}
\eq{\omega_{0,2,2}= \pm {\bf 0.747}289 - {\bf 0.177}543 \imath ,}
where the digits in bold agree with the reference numerical results from \cite{bertiweb} with the rounding in the last bold digit.

\section{Conclusion}\label{sec:conclusion}

In this paper, we studied linear perturbations in asymptotically flat Schwarzschild spacetime, focusing on the small-frequency regime. The local solutions in the near-spatial infinity region are determined in terms of multiple polyexponential integrals \cite{Aminov:2024aan}.
Considering the properties of these solutions under the monodromy transformations, we obtained a partial differential equation on the scattering elements, which fixes their exact dependence on $\log t$. The remaining constant of integration was related to the Seiberg-Witten quantum period via the resummation of infinitely many instantons. This led to an alternative formulation of the scattering amplitudes in terms of the Nekrasov-Shatashvili free energy. We finally studied the poles of the scattering elements and found agreement with the previous results obtained via the Seiberg-Witten theory \cite{Aminov:2020yma}.
Using the small-frequency expansion, we computed fundamental QNM frequencies of the Schwarzschild spacetime for different values of the quantum numbers.
The advantage of our method is that it only requires two Taylor expansions in $t$ as an input:
\eq{\beta_{\ell,s}=\sum_{k\geq 1} \beta_{\ell,s,k} \, t^{2k},\quad 
\varphi_{\ell,s} = \sum_{k\geq 0} \varphi_{\ell,s,k} \,t^{2\,k}.}
As a downside, the convergence properties of these Taylor expansions do not allow one to find results for higher overtones. It would be interesting to see if analogous expansions exist for higher overtone numbers. 

In deriving scattering elements, we used perturbative expansions of multiple polyexponential integrals and polylogarithms, which contain multiple zeta values. Remarkably, in the expansion of the scattering elements themselves, all multiple zeta values cancel, and only ordinary zeta values remain. In part, this could be explained by our result for the inverse scattering elements \eqref{exactGreen}, where the only functions involved are Gamma functions and trigonometric and hyperbolic functions. The less clear part is that the coefficients in the Taylor expansions of $\beta_{\ell,s}$, $\varphi_{\ell,s}$, and $\xi_{\ell,s}$ are rational numbers, except for $\varphi_{0,0}=\log\brc{7/9}$. It would be nice to have clear physical explanation of the latter results.

A more efficient way of computing the coefficients in the Taylor expansions of $\beta_{\ell,s}$ and $\varphi_{\ell,s}$ might exist since they are mostly rational numbers. For example, the two local solutions in the near-spatial infinity region satisfy a second-order partial differential equation similar to (\ref{eq:G_diff}):
\eq{\frac{\partial^2 \psi^{\text{in},\text{out}}_{(\infty)}}{\partial \brc{\log z_L}^2} + t\,
\frac{\partial \psi^{\text{in},\text{out}}_{(\infty)}}{\partial \log z_L} +\brc{\frac{t^2}{4} -\beta^2_{\ell,s}} \psi^{\text{in},\text{out}}_{(\infty)} =0,}
which fixes their exact dependence on $\log z_L$:
\eqlb{eq:psi_logz}{\psi^{\text{in},\text{out}}_{(\infty)} =c^{\text{in},\text{out}}_1 z_L^{\beta_{\ell,s}-t/2} +c^{\text{in},\text{out}}_2 z_L^{-\beta_{\ell,s}-t/2}.}
Here the integration constants are functions of $z_L$: $c_{1,2}^{\text{in},\text{out}}\equiv c_{1,2}^{\text{in},\text{out}}\brc{z_L}$. As one can see, $\beta_{\ell,s}$ plays an important role in the Taylor expansion of (\ref{eq:psi_logz}) around $z_L=0$. Thus, imposing particular constraints on $c_{1,2}^{\text{in},\text{out}}$ might lead to an alternative derivation of $\beta_{\ell,s}$ using the Frobenius or Floquet-type expansions \cite{Frobenius1873, %ASENS_1883_2_12__47_0,
Lisovyy:2021bkm}.

In this work, we did not consider other types of $4$d black holes. However, applying our method to the Kerr or Reissner-Nordstr\"om black holes should be straightforward, as their perturbations are described by the same class of ordinary differential equations of second order with an irregular singularity. In these cases, the local solutions in the small-frequency regime should also be described in terms of multiple polylogarithms and multiple polyexponential integrals. %Another possible direction would be to apply the analytic small-frequency expansion to studying other gravitational quantities, such as greybody factors or Love numbers.

As it was pointed out in the introduction, other perturbative methods have been developed to study QNMs and scattering amplitudes, as the instanton approach \cite{Aminov:2020yma} and the MST method \cite{Mano:1996mf,Mano:1996gn,Mano:1996vt}. We investigated in detail the relation between our method and the instanton approach. In the same way, it would be important to relate it also with the MST method. The latter involves a matching of asymptotic expansions in which the local solutions are expressed as series of hypergeometric functions depending on an additional parameter $\nu$ - the \emph{renormalized
angular momentum}. The coefficients in such series expansions obey a three-term recurrence relation, which permits to express the parameter $\nu$ and the outgoing and incoming amplitudes as expansions in powers of $\epsilon=2\,M\,\omega$ (which equals the instanton parameter that we denoted with $t$ up to a constant factor of $-2\,\imath$).

The expansions provided by MST method were used in \cite{Ivanov:2022qqt,Saketh:2023bul,Ivanov:2024sds} to study the dynamical tidal response of black holes by matching them with the predictions of the point particle effective field theory. This tidal response of black holes is characterized by the so-called dynamical Love numbers.  In particular, it was shown that the wave amplitude ratio factorizes into two parts: the near-zone, which carries information about finite-size effects due to the black hole, and the far-zone, which contains the relativistic post-Minkowskian corrections. This factorization of the scattering elements was then analyzed in \cite{Bautista:2023sdf}, where it was rewritten in the gauge theory language. In this way, the connection between the instanton approach and the MST method was established, and the parameter $\nu$ of the MST method was identified with the gauge modulus $a$. By the same logic, our parameter $\beta$ is also related to the renormalized
angular momentum $\nu$ through (\ref{eq:a_beta_rel}). After establishing the exact form of this relation, it would be essential to derive the corresponding dynamical Love numbers.

Another possible direction would be to study the spectral stability of black hole QNM frequencies, namely evaluating how they move in the complex plane under small perturbations of the differential operator \cite{Nollert:1996rf,Nollert:1998ys}. 
This problem is related to the non-normality of the operators governing the black hole linear perturbations with respect to an energy inner product \cite{Carballo:2024kbk}.
For self-adjoint operators in non-dissipative systems, the spectrum is stable under perturbations \cite{TrefethenEmbree+2005}, meaning that small perturbations of the operator (in some fixed norm) lead to small movements of the operator’s eigenvalues on the
complex plane. In the case of dissipative systems such as black holes, instead, instabilities could arise because of the non-completeness of the set of eigenfunctions.
The response of the Schwarzschild black
hole spectrum to specific potential perturbations was studied in \cite{Jaramillo:2020tuu,Cheung:2021bol,Konoplya:2022pbc}.

%Normal operators are defined by having a complete and orthogonal set of eigenfunctions with respect to a given inner product

\appendix

\section{Results for \texorpdfstring{$\beta$}{}, \texorpdfstring{$\xi$}{}, and \texorpdfstring{$\varphi$}{} expansions}\label{appendix:beta}

The small-frequency approach described in Sections \ref{sec:RW} and \ref{sec:wavesol} provides us with the following 
perturbative results for $\beta_{\ell,s}$, $\xi_{\ell,s}$, and $\varphi_{\ell,s}$ with $\ell=s=0,1,2$. When $\beta_{\ell,s}$ and $\xi_{\ell,s}$ are determined up to order $t^K$, the corresponding expansions of $\varphi_{\ell,s}$ can be computed up to order $t^{K-2\,\ell-2}$. In case $\ell=s=0$, the wave functions in the near-spatial infinity region were computed up to order $t^{16}$, which fixes the first $14$ orders of $\beta_{0,0}$ and $\xi_{0,0}$ and $12$ orders of $\varphi_{0,0}$:
\begin{equation}
\begin{aligned}
\beta_{0,0}=\frac{7}{24}\, t^2-\frac{9449}{120\,960}\,t^4+\frac{102\,270\,817 }{2\,133\,734\,400}\,t^6 -\frac{4\,988\,909\,608\,861}{150\,556\,299\,264\,000}\,t^8 +\frac{72\,237\,319\,625\,071\,987}{2\,655\,813\,119\,016\,960\,000}\,t^{10}& \\
-\frac{2\,008\,359\,560\,158\,182\,591\,511}{86\,646\,394\,825\,172\,582\,400\,000} \, t^{12} + \frac{202\,956\,264\,764\,788\,667\,222\,561\,313\,859}{9\,656\,699\,112\,995\,968\,225\,640\,448\,000\,000} \,t^{14} + \mathcal{O}\left(t^{16}\right),&
\end{aligned}
\end{equation}
\begin{equation}
\begin{aligned}
\xi_{0,0}(t)=1-\frac{5}{6}\,t+\frac{25}{72}\,t^2-\frac{23}{315}\,t^3+\frac{41}{72\,576}\,t^4+\frac{225\,487}{17\,781\,120}\,t^5-\frac{51\,838\,301}{6\,401\,203\,200}\,t^6+\frac{9\,481\,089\,257}{1\,568\,294\,784\,000}\, t^7&\\
-\frac{284\,742\,073\,739}{90\,333\,779\,558\,400}\,t^8+\frac{722\,761\,773\,679\,487}{304\,311\,919\,887\,360\,000}\,t^9-\frac{108\,778\,841\,322\,632\,893 }{87\,641\,832\,927\,559\,680\,000}\,t^{10}&\\
+\frac{1\,711\,007\,301\,901\,126\,757\,149 }{1\,754\,589\,495\,209\,744\,793\,600\,000}\,t^{11}-\frac{10\,726\,170\,068\,189\,907\,710\,593}{21\,055\,073\,942\,516\,937\,523\,200\,000}\,t^{12}&\\
+\frac{4\,383\,394\,423\,080\,102\,339\,340\,164\,317}{10\,622\,369\,024\,295\,565\,048\,204\,492\,800\,000}\,t^{13}&\\
-\frac{68\,395\,977\,167\,815\,587\,016\,740\,406\,979}{318\,671\,070\,728\,866\,951\,446\,134\,784\,000\,000}\,t^{14}+ \mathcal{O}\left(t^{15}\right),&\\
\end{aligned}
\end{equation}
\begin{equation}
\begin{aligned}
\varphi_{0,0}&=\log\brc{ \frac{7}{9}}-\frac{8587}{70\,560}\,t^2 +\frac{59\,423\,233}{995\,742\,720}\,t^4-\frac{3\,034\,619\,927\,027}{131\,736\,761\,856\,000}\,t^6+\frac{2\,636\,632\,572\,686\,279\,887}{136\,331\,740\,109\,537\,280\,000}\,t^8 \\
&-\frac{6\,348\,255\,806\,061\,211\,753\,575\,743}{429\,874\,426\,326\,387\,474\,432\,000\,000}\,t^{10}+\frac{120\,127\,293\,342\,168\,835\,533\,078\,304\,741}{9\,463\,565\,130\,736\,048\,861\,127\,639\,040\,000}\,t^{12}+\mathcal{O}\left(t^{14}\right).
\end{aligned}
\end{equation}
For $\ell=s=1$, we computed the near-infinity local wave functions up to order $t^{14}$, which gives
\begin{equation}
\begin{aligned}
\beta_{1,1}=\frac{47}{240}\,t^2-\frac{43\,908\,007}{1\,137\,024\,000}\,t^4+\frac{1\,897\,955\,762\,232\,049}{126\,588\,975\,206\,400\,000}\,t^6-\frac{916\,976\,100\,036\,495\,015\,111\,773}{124\,023\,735\,704\,345\,444\,352\,000\,000}\,t^8 &\\
+\frac{5\,706\,721\,543\,769\,515\,470\,350\,083\,430\,384\,773}{1\,410\,389\,488\,815\,788\,497\,680\,728\,064\,000\,000\,000} \,t^{10} &\\
- \frac{106\,995\,634\,360\,703\,511\,437\,460\,300\,615\,557\,539\,248\,229\,319}{44\,908\,789\,408\,918\,140\,501\,620\,712\,312\,039\,014\,400\,000\,000\,000}\,t^{12}
+\mathcal{O}\left(t^{14}\right),&
\end{aligned}
\end{equation}
\begin{equation}
\begin{aligned}
\xi_{1,1}(t)=-1+\frac{5}{4}\,t -\frac{25}{32}\, t^2 + \frac{1471}{4512}\, t^3 - \frac{29\,555}{288\,768}\,t^4 +\frac{26\,305\,804\,327}{1\,004\,674\,406\,400}\,t^5 -\frac{38\,025\,119\,711}{6\,429\,916\,200\,960}\,t^6 &\\
+\frac{152\,296\,572\,400\,211\,831 }{111\,854\,018\,492\,375\,040\,000}\,t^7 -\frac{1\,089\,548\,738\,109\,409\,027 }{2\,863\,462\,873\,404\,801\,024\,000}\,t^8&\\
+\frac{2\,259\,394\,074\,262\,863\,197\,636\,203}{15\,655\,338\,981\,194\,233\,518\,489\,600\,000}\,t^9-\frac{1\,293\,517\,008\,380\,421\,728\,073\,421 }{20\,038\,833\,895\,928\,618\,903\,666\,688\,000}\,t^{10}&\\
+\frac{1\,549\,762\,537\,865\,582\,971\,981\,233\,948\,038\,333}{48\,554\,031\,908\,479\,118\,826\,650\,311\,065\,600\,000\,000}\,t^{11}&\\
-\frac{6\,224\,927\,205\,898\,953\,592\,055\,383\,850\,794\,531}{395\,494\,659\,909\,066\,276\,987\,987\,988\,316\,160\,000
\,000}\,t^{12}+ \mathcal{O}\left(t^{13}\right),&\\
\end{aligned}
\end{equation}
\begin{equation}
\begin{aligned}
\varphi_{1,1}=\frac{895\,597}{7\,068\,800}\,t^2-\frac{455\,691\,732\,736\,543}{4\,407\,171\,729\,408\,000}\,t^4+\frac{74\,327\,495\,711\,146\,205\,449\,777}{1\,226\,665\,736\,133\,046\,272\,000\,000}\,t^6& \\
-\frac{29\,832\,726\,638\,753\,503\,996\,423\,316\,486\,442\,073}{793\,193\,404\,821\,187\,035\,447\,794\,073\,600\,000\,000}\,t^8+\mathcal{O}\left(t^{10}\right)&.
\end{aligned}
\end{equation}
For $\ell=s=2$, we also computed the near-infinity local wave functions up to order $t^{14}$ and got the following results for $\beta_{2,2}$, $\xi_{2,2}$, and $\varphi_{2,2}$:
\begin{equation}
\begin{aligned}
\beta_{2,2}=\frac{107}{840}\,t^2-\frac{1\,695\,233}{148\,176\,000}\,t^4+\frac{76\,720\,109\,901\,233}{30\,764\,716\,012\,800\,000}\,t^6-\frac{71\,638\,806\,585\,865\,707\,261\,481}{99\,644\,321\,084\,605\,000\,704\,000\,000}\,t^8&\\
+\frac{270\,360\,664\,939\,833\,821\,554\,899\,493\,653\,643}{1\,152\,641\,264\,228\,149\,083\,523\,559\,424\,000\,000\,000}\,t^{10}&\\
-\frac{25\,911\,378\,819\,560\,727\,799\,984\,792\,720\,318\,253\,742\,297\,427}{317\,331\,168\,450\,503\,888\,940\,177\,332\,763\,456\,307\,200\,000\,000\,000}\,t^{12}+\mathcal{O}\left(t^{14}\right),&
\end{aligned}
\end{equation}
\begin{equation}
\begin{aligned}
\xi_{2,2}(t)=1-\frac{5}{3}\,t +\frac{25}{18}\, t^2 - \frac{73}{96}\, t^3 + \frac{785}{2592}\,t^4 -\frac{1\,007\,354\,009}{10\,871\,884\,800}\,t^5 +\frac{896\,782\,589}{39\,138\,785\,280}\,t^6  &\\
-\frac{6\,050\,546\,248\,023\,481}{1\,231\,227\,907\,338\,240\,000}\,t^7+\frac{39\,569\,841\,898\,687}{41\,040\,930\,244\,608\,000}\,t^8&\\
-\frac{49\,266\,134\,378\,785\,551\,042\,377}{306\,757\,182\,671\,647\,123\,046\,400\,000}\,t^9 \frac{6\,918\,422\,754\,413\,573\,300\,251}{368\,108\,619\,205\,976\,547\,655\,680\,000}\,t^{10} &\\
-\frac{8\,229\,146\,383\,510\,495\,333\,773\,316\,804\,837}{3\,548\,430\,888\,956\,507\,708\,078\,122\,598\,400\,000\,000}\,t^{11}&\\
+\frac{1\,543\,339\,979\,694\,704\,523\,189\,885\,578\,017}{2\,129\,058\,533\,373\,904\,624\,846\,873\,559\,040\,000\,000}\,t^{12}+ \mathcal{O}\left(t^{13}\right),&\\
\end{aligned}
\end{equation}
\begin{equation}
\varphi_{2,2}=\frac{125}{1568}\,t^2-\frac{53\,950\,959\,337}{2\,280\,051\,680\,256}\,t^4+\frac{29\,227\,746\,558\,029\,477\,261}{3\,905\,473\,390\,495\,507\,353\,600}\,t^6+\mathcal{O}\left(t^{8}\right).
\end{equation}

It is possible to determine the generic expressions for $\beta_{\ell,s,k}$, $\ell\ge k$ and $\varphi_{\ell,s,k}$, $\ell>k$ using the instanton part of the NS free energy $\mathcal{F}^{(\text{NS})}_{\text{inst}}$  with $N_f=3$. We start by inverting the Matone relation to determine the modulus $a$ as a function of $\omega$, $\ell$, $s$, and the instanton parameter $\Lambda$:
\eqlb{aMatonerel}{a^2=2\,\omega^2 -\brc{\ell+\frac{1}{2}}^2 +\Lambda\, \frac{\partial \mathcal{F}^{(\text{NS})}_{\text{inst}}}{\partial \Lambda}.}
In general, when getting the small-frequency expansion of $a$, infinitely many orders in $\Lambda$ can contribute to the same order in $t$ or $\omega$. However, when $\ell\ge k\ge1$, the first $2\,k$ instantons are enough to compute the $t$-expansions of $a$ and $\beta_{\ell,s}$ up to order $t^{2k}$.
The perturbative expansion of $\varphi_{\ell,s}$ can be determined by taking the $a$-derivative of $\mathcal{F}^{(\text{NS})}_{\text{inst}}$ and substituting the instanton expansion of $a$ obtained earlier by inverting the Matone relation. Unless $\ell>k$, infinitely many orders in $\Lambda$ contribute to the same order $t^k$ of $\varphi_{\ell,s}$. For $\ell>k$, we use the first $2\,k$ instantons to obtain the generic expressions of $\varphi_{\ell,s,k}$.

Below are the results for $\beta_{\ell,s,1}$, $\ell\geq 1$ and $\beta_{\ell,s,2}$, $\ell \geq 2$. To simplify the expressions, we introduce the notation $L\equiv \ell\brc{\ell+1}$:
\begin{equation}
\ell\geq 1:\quad \beta_{\ell,s,1}=\frac{1}{2\brc{2\,\ell-1}_5}\brc{15 \,L^2 -11\, L + 6\,s^2 \left(L-1\right) +3\, s^4},
\end{equation}
\begin{equation}
\begin{aligned}
\ell\geq 2:\quad \beta_{\ell,s,2}=-&\frac{2}{\brc{\brc{2\,\ell-1}_5}^2 \brc{2\,\ell-3}_9 }\Bigl\{9\,s^8 \brc{L-2} \brc{560\, L^3-840\, L^2+63\, L+45}\\
&+4\,s^2\,L\brc{5040\, L^6-39\,480\, L^5+106\,015\, L^4-124\,514\, L^3+58\,737\, L^2-3528\, L-2970}\\
&+2\, s^4 \brc{8400\, L^6-65\,800\, L^5+171\,689\, L^4-152\,670\, L^3+40\,113\, L^2+4752\, L-1620}\\
&+L^2\, \brc{L-2} \brc{18\,480\, L^5-105\,000 \, L^4+155\,295\, L^3-82\,625 \, L^2+8733 \, L+3240}\\
&+4\, s^6 \brc{2800\, L^5-19\,880\, L^4+32\,907\, L^3-16\,731\, L^2-81\, L+810}\Bigr\},
\end{aligned}
\end{equation}
where $\brc{q}_n$ is the rising Pochhammer symbol.

Similarly, we have for $\varphi_{\ell,s,1}$ and  $\varphi_{\ell,s,2}$:
\begin{equation}
\begin{aligned}
\ell>1\colon\quad\varphi_{\ell,s,1}=\frac{(2 \ell+1)^3}{2\brc{\brc{2\,\ell-1}_5}^2} \left[L^2+4\,s^2 \brc{14\,L^2-24\, L+9}+3\,s^4 \brc{8\, L-3}\right],
\end{aligned}
\end{equation}
\begin{equation}
\begin{aligned}
\ell>2\colon\quad\varphi_{\ell,s,2}=\,&-\frac{4 (2\,\ell+1)^3}{3\left((2\, \ell-3)_9\right){}^2 \left((2\,\ell-1)_5\right){}^2}\Bigl\{3\brc{L-2}^2 L^2\,\bigl(622\,848\, L^7-6\,212\,544\, L^6+21\,712\,272\, L^5\\
&-32\,554\,372\, L^4+22\,795\,266\, L^3-5\,708\,997\, L^2-1\,088\,640\, L+583\,200\bigr)+4\,L\,s^2\,\bigl(1\,061\,376\, L^9\\
&-15\,044\,160\, L^8+86\,481\,808\, L^7-260\,933\,804\, L^6+444\,628\,113\, L^5-420\,000\,264\, L^4+187\,434\,000\, L^3\\
&-5\,375\,808\, L^2-25\,048\,440\, L+6\,415\,200\bigr)+4\,s^4\,\bigl(1\,018\,752\, L^9-14\,267\,744\, L^8+79\,516\,264\, L^7\\
&-221\,705\,538\, L^6+322\,081\,092\, L^5-232\,063\,569\, L^4+61\,460\,370\, L^3+15\,646\,770\, L^2-12\,830\,400\, L\\
&+2\,187\,000\bigr)+4\,s^6\,\bigl(735\,232\, L^8-9\,344\,576\, L^7+43\,334\,736\, L^6-93\,552\,300\, L^5+98\,731\,791\, L^4\\
&-43\,634\,700\, L^3-1\,363\,716\, L^2+6\,765\,120\, L-1\,749\,600\bigr)+135\brc{L-2}^2 s^8\,\bigl(8448\, L^5-41\,152\, L^4\\
&+49\,488\, L^3-5076\, L^2-5130\, L+2025\bigr) \Bigr\}.
\end{aligned}
\end{equation}

As mentioned earlier, the distinction between the cases in which a finite or an infinite number of instantons contribute to the same order in $t$ arises from the presence of the poles at integer values of $\ell$ in the NS function.  We remark that the resummation of infinitely many instantons in the $t$-expansion differs from the apparent resummation in the pure instanton approach. There, this resummation may be avoided by analytically continuing $\ell$ to be a generic complex number, and taking the limit $\ell\to\mathbb{Z}_{\ge 0}$ only in the final step of the computation. This procedure is discussed in \cite{Bautista:2023sdf}, where the authors show that the \emph{fixed-$\ell$ prescription} and the \emph{generic-$\ell$ prescription} lead to the same answer for the computation of the scattering phase-shift in Kerr spacetime. The same reasoning also applies when computing the coefficients of the QNM frequencies of Schwarzschild-de Sitter spacetime in 4 dimensions in the Taylor series expansion around $R_h=0$ ($R_h$ being the radius of the event horizon) as discussed in footnote 9 in \cite{Arnaudo:2024rhv}.

 \bibliographystyle{JHEP}
 \bibliography{biblio}

\end{document}